# Band-Edge Carrier Trapping Limits Light Emission in WSe$_2$

*Juri G. Crimmann,[†] Sander J. W. Vonk,[†] Yannik M. Glauser,[†]*
*Gabriel Nagamine,[†,§] and David J. Norris[\*,†]*

[†]Optical Materials Engineering Laboratory, Department of Mechanical and Process Engineering, ETH Zurich, 8092 Zurich, Switzerland

[§]Institute of Applied Physics, University of Bern, 3012 Bern, Switzerland

ABSTRACT: Monolayers of transition metal dichalcogenides (TMDs) exhibit bright photoluminescence, a desirable property for light-emitting diodes and single-photon emitters. Because the emission intensity is heavily influenced by factors such as defect density and oxidation, it is critical to understand how they affect photoluminescence efficiency. However, due to the time-consuming process of identifying individual monolayers, studies of high-quality exfoliated TMDs have been limited to only a few samples. Here, we present an investigation of excited-state lifetimes and spectra for over 200 WSe$_2$ exfoliated monolayers at room temperature. We find a linear correlation between photoluminescence lifetime and intensity across hundreds of monolayers and within individual monolayers. Results from intentional photooxidation experiments indicate that this correlation is due to photoinduced band-edge carrier traps, which introduce a nonradiative decay pathway that competes with exciton emission. Our work highlights the importance of controlling such traps, as they are the primary limitation of bright photoluminescence.

Transition metal dichalcogenides (TMDs) such as WSe$_2$, MoSe$_2$, and MoS$_2$ are layered semiconductors bound by van-der-Waals forces.[1] Unlike their bulk counterparts, monolayers of these TMDs are direct band gap semiconductors,[2] leading to bright photoluminescence (PL).[3,4] Consequently, monolayer TMDs are promising candidates as the active material for light-emitting diodes,[5] lasers,[6] and single-photon emitters.[7] Since they are atomically thin, quantum confinement and reduced dielectric screening result in strong Coulombic attraction between electrons and holes.[8,9] Hence, binding energies for electron–hole pairs (excitons) of several hundred meV have been reported,[10] and even charged excitons (trions) have been observed.[11] These excited states can recombine radiatively through light emission.[12]

However, nonradiative recombination can also occur in TMDs.[13,14] For example, trions,[15] trapped excitons, and individually trapped electrons or holes can participate in nonradiative relaxation processes such as exciton–phonon scattering, inelastic Auger scattering, or defect-assisted recombination.[12,15] Dark-exciton states have also been observed, which quench light emission at low temperatures.[16] While all of the above mechanisms can reduce the photoluminescence quantum yields (PLQYs) of TMD monolayers, not all of them are equally important. Therefore, it is essential to have a good understanding of the decay mechanisms in monolayer TMDs to bring out their full potential.

Until now, studies have been challenged by the process of obtaining high-quality monolayers. They are typically prepared through time-intensive mechanical exfoliation, followed by manual search in a brightfield (BF) microscope.[2,17] In practice, researchers spend significant time preparing TMD samples. This process has limited previous studies to a small number of investigated monolayers,[12,18] which can lead to potentially inaccurate conclusions about the dominant nonradiative decay pathways. However, automated PL identification enables high-throughput characterization of exfoliated monolayers.[19] Such a technique allows the investigation of decay mechanisms beyond a single sample, providing statistically relevant data on the nonradiative decay process.

Here, we study the room-temperature emission from WSe$_2$ monolayers by combining automated PL identification with automated PL characterization of TMD monolayers. More specifically, we correlate excited-state lifetimes and brightness of individual WSe$_2$ flakes. In total, we characterize

>200 exfoliated WSe$_2$ monolayers of two different bulk crystals. In addition, we spatially map lifetimes and PL intensities within a single monolayer. We find a linear correlation between intensity and excited-state lifetime across batches and within monolayers, meaning that bright PL correlates with slow lifetime and dim PL with fast lifetime. This observation can be explained by trap states, which quench PL by introducing nonradiative decay pathways. From systematic photobleaching experiments on a single WSe$_2$ monolayer, we conclude that the traps are photoinduced. Our observations are also consistent with band-edge carrier trapping that reduces the radiative recombination.

We begin by studying two WSe$_2$ monolayers of two different bulk crystals from the same supplier. They were prepared via mechanical exfoliation (see Methods in the Supporting Information) on a polydimethylsiloxane (PDMS) substrate supported by a glass microscope slide. The monolayers were later identified using a BF microscope (see Figure S1). We acquired atomic force microscopy images (AFM, see Figure S2) and Raman measurements[3,20] (Figure S3), confirming that they are single sheets of WSe$_2$. Notably, these two monolayers appear similar when investigated using these methods.

However, optical characterization reveals significant differences in PL brightness between the two flakes (Figure 1a,b) when images are recorded at equal excitation powers. The two monolayers show different PL intensities overall; one is dim and the other is bright. Furthermore, the PL is inhomogeneous across each flake (grainy texture, discussed below), and the center of each flake is brighter than its edges. We exclude local strain as the cause of these inhomogeneities, as the AFM (Figure S2) and PL images do not show the same spatial variations.

To investigate the origin of these differences in the PL intensity, we measured time-resolved PL (Figure 1c) and emission spectra (Figure 1d).[12,21] While the spectra of the two monolayers differ in intensity, their line shapes are the same (see inset). However, time-resolved PL measurements show more pronounced variations, with the dim flake (blue) having a significantly shorter excited-state lifetime than the bright flake (red). Although it was suggested that higher defect densities can lead to lower PL brightness,[22-24] the influence of defects on excited-state lifetimes remains challenging to quantify and isolate.[15]

We investigated the different contributions to the emission spectra and decay curves by performing power-dependent measurements on the bright flake (Figure S4). For this, we changed the excitation power in discrete steps from 0.002 µW to 0.030 µW. For every excitation power, we acquired PL spectra (Figure S4a) and decay curves (Figure S4b). The spectra change in intensity, but their line shapes remain the same (Figure S4a, inset). We fitted the decay curves using a double-exponential, convoluted with a Gaussian instrument response function. The decay curves consist of a short-lifetime component ($\tau_{short}$ = 0.8 ns) and a long-lifetime component ($\tau_{long}$ = 3.2 ns). In Figure 1e and Figure S4c,d, we plot the amplitudes versus excitation power. For low excitation powers, the amplitudes of the long component follow a linear power increase, which is consistent with the formation of excitons. However, the amplitudes of the short component exhibit a quadratic power scaling, in line with the formation of biexcitons or charged biexcitons.[12,21,25-29] The biexcitonic emission has a shorter decay due to its faster radiative lifetime and its additional nonradiative Auger recombination.[15,21,27,30]

To address whether the variations in brightness and excited-state lifetimes observed above are representative for the entire batch, we automated the identification[19] and characterization of our TMD samples. We acquired PL lifetimes and spectra for >200 WSe$_2$ monolayers exfoliated from the two different bulk crystals (batches). Figure 2a shows the distribution of the PL intensities. To compare the intensities of the two batches (blue and red), each batch was normalized to compensate for its slightly different excitation fluence. Figure 2b shows the distribution of the average lifetimes, which we use to reduce the multiexponential decay dynamics to a single number[31] (see Supporting Information). The blue batch has an average lifetime of $\tau_{avg}$ = (0.66 ± 0.08) ns (mean ± standard deviation), and the red batch of $\tau_{avg}$ = (2.3 ± 0.5) ns. Figure 2c shows the intensity versus lifetime distributions for all of our samples. The blue and red data show a positive correlation between intensity and lifetime, both within one batch and between them. In short, we observe that in WSe$_2$ monolayers bright PL correlates with a long average lifetime (discussed further below). We conclude that the dim and bright monolayers in Figure 1 are indeed representative, belonging to a dim and bright batch (blue and red, respectively).

Our method also allows us to simultaneously perform high-throughput spectral measurements to investigate the origin of the observed differences in PL. Figure 3a shows two selected spectra of monolayers from each batch (blue and red). The two emission features in TMD monolayers have previously been attributed to excitons and trions.[29,32,33] Unlike in the time-resolved PL measurements, biexciton and/or charged biexciton emission is not visible in the PL spectra. At room temperature, thermal broadening of the strong exciton and trion peaks likely obscures the weaker biexcitonic emission.[34] We fit the spectra with the sum of two Gaussians (see Supporting Information). For this specific pair of samples, we observe that the trion and exciton emissions from the blue batch are shifted to lower energy compared to those from the red batch.

To explore such differences more systematically, we fitted emission spectra from hundreds of monolayers and analyzed the resulting peak positions. Figure 3b,c shows the distribution of peak emission energies of the exciton and trion emissions. The blue batch has average energies for the exciton and trion peaks of $E_{avg}^X = (1.675 \pm 0.002)$ eV and $E_{avg}^T = (1.656 \pm 0.004)$ eV, where the second number represents the standard deviation in each distribution. Similarly, the values for the red batch are $E_{avg}^X = (1.676 \pm 0.002)$ eV and $E_{avg}^T = (1.658 \pm 0.002)$ eV. On average, the exciton and trion emission peaks from the blue batch are shifted by ~1 meV to lower energy compared to those of the red batch. Notably, this shift is on the same order of magnitude as the spread of peak positions within each batch. To analyze whether these differences are statistically significant, we performed t-tests. Given the large sample sizes ($N_{blue} = 134$, $N_{red} = 100$), the shifts in average peak positions are highly significant [exciton: $t(232) = 3.78$, $p < 0.001$, trion: $t(232) = 5.01$, $p < 0.0001$]. Notably, these small average shifts only become apparent through the statistical analysis of hundreds of monolayers. Indeed, the data for only two samples shown in Figure 3a would suggest much larger shifts.

The shift in emission between the two batches can yield a clue about variations in their PL intensity. Previously, it was reported that regions of increased defect density and oxidation in a single CVD-grown monolayer show a lower energy in the PL spectrum.[35] Similarly, exfoliated samples also exhibit such a shift caused by degradation over time.[36] Increased oxidation, confirmed by X-ray photoelectron spectroscopy (XPS), was connected to reduced PL intensities and shifted emission to lower energies.[36-40] We observe the same behavior for our two batches, suggesting that the blue batch has more defects and/or is more oxidized than the red batch.

We now return to the positive correlation between PL intensity and average lifetime observed in Figures 1 and 2. To find its origin, we focus on a single monolayer. More specifically, we spatially map the intensity (Figure 4a) and average lifetime (Figure 4b) of the bright flake in Figure 1 by confocal scanning. Again, as in Figure 1, we observe dim edges compared to the bright interior. Similarly, the average-lifetime map reveals a long-lifetime interior and short-lifetime edges (Figure 4b). Figure 4c shows two example decay curves, one from the center and one from the edge. In Figure 4d,e, we plot how the intensities and average lifetimes change with the distance to the edge. These graphs again demonstrate that the PL intensities and average lifetimes follow the same trend. We investigate their correlation in Figure 4f, which plots intensities against average lifetimes, with each pixel color coded by its shortest distance to the edge of the monolayer. The observed linear correlation (dashed line, see Table S1) in Figure 4f holds both at the edges of the monolayer (dark points) as well as in the center (bright points). This suggests that the same underlying mechanism is responsible for both the gradient towards the edges and the inhomogeneous emission within the monolayer (grainy texture). The few outliers (~1% of the total measurements) originate only from the edges of the monolayer, where the relatively high background (due to a low emission intensity) leads to longer average lifetimes during the fitting procedure. We note that the linear behavior cannot be explained by variations in local charging as the spectra remain very similar over the entire flake (Figure S5).

The question arises: what can cause a reduction in PL intensity while simultaneously shortening the lifetime? One possible explanation is photoinduced oxidation, which was previously observed to be strongest at the edges of CVD-grown TMD monolayers.[35,38] Prior studies have also connected laser irradiation and oxidation, providing a route to locally oxidize a WSe$_2$ monolayer.[40-42] To understand the role of oxidation in our exfoliated TMDs, we systematically bleached a WSe$_2$ monolayer using a 405-nm laser. We exposed a monolayer under ambient conditions at 6 different spots to a defocused laser with powers between 1–6 mW for 10 s. Afterwards, the PL image (Figure 5a) clearly reveals the effect of the laser irradiation. The PL signal decreases with increasing laser power. This change did not recover (at least over several weeks). Despite the strong bleaching observed, no physical damage can be seen in the BF image of Figure 5a.

After bleaching we then waited several hours before collecting further data from the monolayer, ensuring that the sample was not temporally charged or ionized. In Figure 5b,c, we compare PL lifetimes and spectra from the center of the bleached spots with those from an unexposed region. Excitation powers (0.01 µW) and collection conditions were constant for all measurements. We observe no changes in the PL line shapes after photobleaching (Figure 5c and Figure S6). However, the excited-state lifetimes are severely shortened, and the intensities decrease with increasing photobleaching (Figure 5b). For each decay curve, we plotted the intensity versus average lifetime in Figure 5d. Again, the data follow a linear trend. This is consistent with the positive correlations observed above in Figures 2c and 4f. Hence, we conclude that oxidation, induced by either photoexcitation or other means, is the primary cause of the reduced PL intensities and shorter lifetimes. This effect is particularly pronounced at the monolayer edges (see Figure 4).

Lastly, we wish to understand how oxidation modifies the relaxation pathways within WSe$_2$ monolayers and leads to the observed reduction in PL intensities and excited-state lifetimes. In general, the emission intensity for excitons is given by

$$I \propto \eta \, N_0 = \frac{k_\text{r}}{k_\text{r} + k_\text{nr}} N_0 \qquad (1)$$

with $\eta$ the PL quantum yield, $k_\text{r}$ and $k_\text{nr}$ the radiative and nonradiative decay rates of the band-edge exciton, and $N_0$ the number of band-edge excitons generated per unit time. Since the excited-state lifetime is given by $\tau = 1/(k_\text{r} + k_\text{nr})$, the emission intensity can also be expressed as

$$I \propto (k_\text{r} N_0)\tau \qquad (2)$$

From this, a linear relationship between the PL intensity and lifetime emerges, provided $k_\text{r}$ and $N_0$ are constant.[43,44]

Different nonradiative decay mechanisms can influence the proportionality constant $k_\text{r}N_0$ in equation 2. First, **charging** might change both $k_\text{r}$ and $k_\text{nr}$, leading to different excited-state lifetimes and intensities. However, the spectral line shape (see Figure 5c and Figure S6) remains unchanged, indicating that the exciton/trion ratio is unaffected by the photobleaching. This eliminates charging as a possible mechanism. Second, **hot-carrier trapping** would introduce a smaller $N_0$ because charge carriers would be lost in the process of cooling (see Figure 5e).[14] This would decrease the PL intensity without affecting the excited-state lifetime and would place the resulting data points in Figure 5d below the blue dashed line connecting the origin and the unbleached reference point ($\tau_0, I_0$). Our measurements in Figure 5d, however, do not follow this trend, ruling out hot-carrier trapping. Third, **band-edge carrier trapping** would introduce additional quenching (Figure 5e), where band-edge excitations trap and decay with increasing nonradiative rate $k_{nr}$.[13,45] In this case, both PL intensities and lifetimes would decrease, and the data would follow a linear trend in plots of intensity versus lifetime, as $k_r N_0$ in equation 2 would be unaffected. This is consistent with our experimental data. Therefore, we conclude that band-edge carrier trapping is the dominant nonradiative process.

While photobleaching reduced the overall emission intensity, the spectral line shapes remained unchanged. This observation can be explained by the fast exciton–trion and trion–exciton conversion times in TMD monolayers, which are on the picosecond timescale and orders of magnitude faster than the observed nanosecond lifetimes.[46,47] Consequently, increased exciton trapping reduces the population of both species equally.

The linear correlation between PL intensities and excited-state lifetimes is observed not only in our intentional photobleaching experiments but also when we investigated a single monolayer (Figure 4f) or many monolayers (Figure 2c). Hence, we conclude that trap-assisted nonradiative recombination competes with radiative recombination in all of our samples. These traps may result from photoinduced oxidation as in the bleaching experiment. In contrast, the traps responsible for the reduced PL in the dim batch (Figure 2) are most likely induced by intrinsic crystal defects. Our findings highlight the importance of preventing oxidation in TMD monolayers, which can be achieved by working in inert atmosphere or encapsulating with hexagonal boron nitride[36] or polymer films.[37] Additionally, chemical treatments can improve optoelectronic properties of TMD monolayers.[48]

In conclusion, we have studied PL lifetimes and spectra of >200 WSe$_2$ monolayers. We observe a linear correlation between the average lifetime and PL intensity within a single monolayer, across many monolayers from different bulk crystals, and for a monolayer that is intentionally photobleached. We explain the correlation through band-edge carrier trapping due to photoinduced traps, which introduce a nonradiative decay pathway that competes with radiative decay. Our work indicates that band-edge carrier trapping is the primary limitation for obtaining bright PL in WSe$_2$ TMDs, showing the importance of preventing the formation of such traps. Chemical treatments potentially offer a promising route for obtaining bright monolayers by reducing the density of photoinduced traps.

## ASSOCIATED CONTENT

**Supporting Information**

The Supporting Information is available free of charge at https://pubs.acs.org.

S1. Methods: sample preparation, photoluminescence measurements, instrument response function, Raman measurements, atomic force microscopy measurements, average-lifetime analysis, data-fitting procedures; S2. Supplementary table: fit function and fitting parameters; S3. Supplementary figures: brightfield images, atomic force microscopy images, Raman spectra, power-dependent photoluminescence measurements, photoluminescence spectra maps, photoluminescence spectra of the photobleached monolayer, instrument response function.

## AUTHOR INFORMATION

**Corresponding Author**

*Email: dnorris@ethz.ch

**ORCID**

Juri G. Crimmann: 0000-0002-0367-5172

Sander J. W. Vonk: 0000-0002-4650-9473

Yannik M. Glauser: 0000-0002-5362-0102

Gabriel Nagamine: 0000-0002-4830-7357

David J. Norris: 0000-0002-3765-0678

**Note**

The authors declare no competing financial interest.

ACKNOWLEDGMENTS

This project was funded by the Swiss National Science Foundation (SNSF) under Award No. 200021-232257. We thank N. Lassaline for stimulating discussions and V. Papp for technical assistance.

# Figures

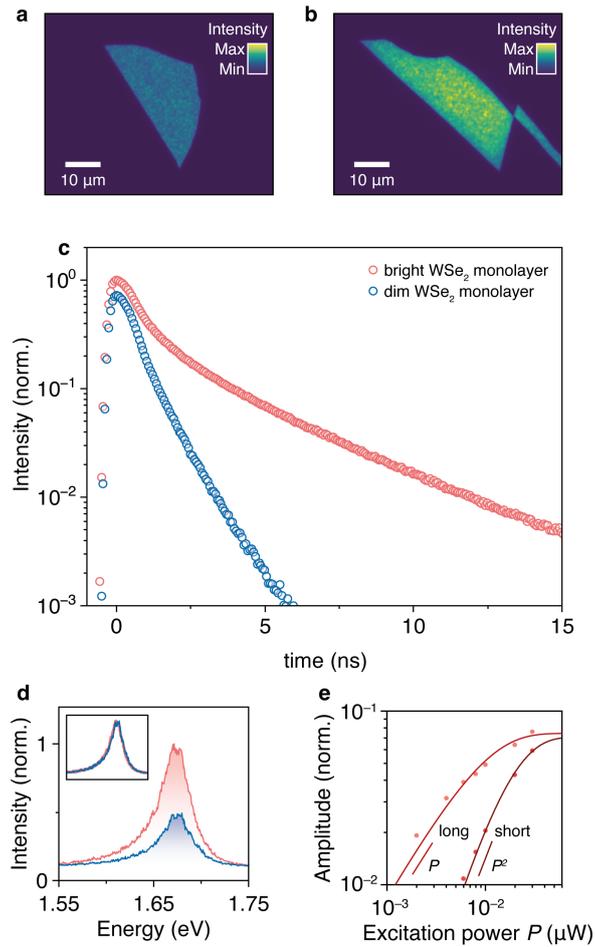

**Figure 1.** Comparing WSe$_2$ monolayers from different bulk crystals. (a,b) Room-temperature photoluminescence images of two WSe$_2$ monolayers from different bulk crystals on PDMS. Images are acquired with the same acquisition settings using a white lamp at equal excitation powers, revealing that one flake is dim and one is bright. (c) Time-resolved PL measurements and (d) emission spectra of the dim monolayer (panel a) in blue, and the bright monolayer (panel b) in red. The inset in d shows the normalized emission spectra. (e) Power-dependent photoluminescence measurements of the bright monolayer of panel b. The extracted amplitudes of the long- and short-lifetime components (points) are plotted against excitation power. Solid lines: expected scaled power-dependent amplitude taking into account the absorption statistics of excitons and biexcitons (or charged biexcitons), the initial rise is linear (exciton, light red) or quadratic (biexciton or charged biexciton, dark red). The fit functions are discussed in more detail in the Supporting Information. The individual decay graphs and emission spectra are shown in Figure S4.

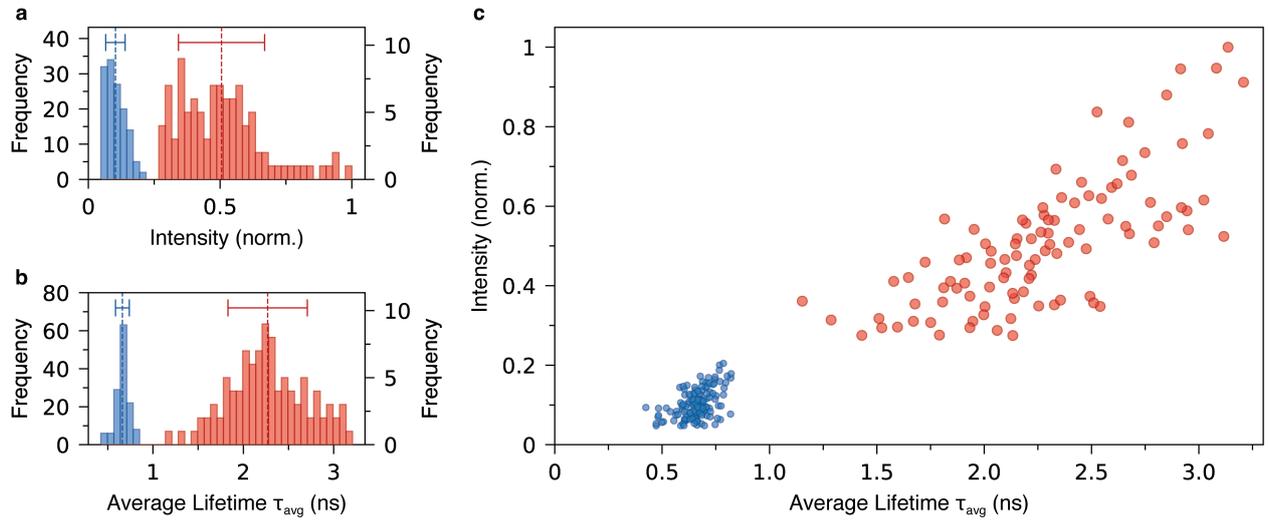

**Figure 2.** Statistical analysis of >200 WSe$_2$ monolayers exfoliated from different bulk crystals (dim and bright batch). (a) Distribution of the normalized PL intensity. To account for slightly different excitation fluences, intensities of each batch were normalized by the respective fluence. (b) Average-lifetime distribution of the dim (blue) and bright (red) batches. The vertical dashed lines and error bars in panels a and b indicate the mean and standard deviation of the distributions. (c) Scatter plot of the normalized PL intensity versus average lifetime.

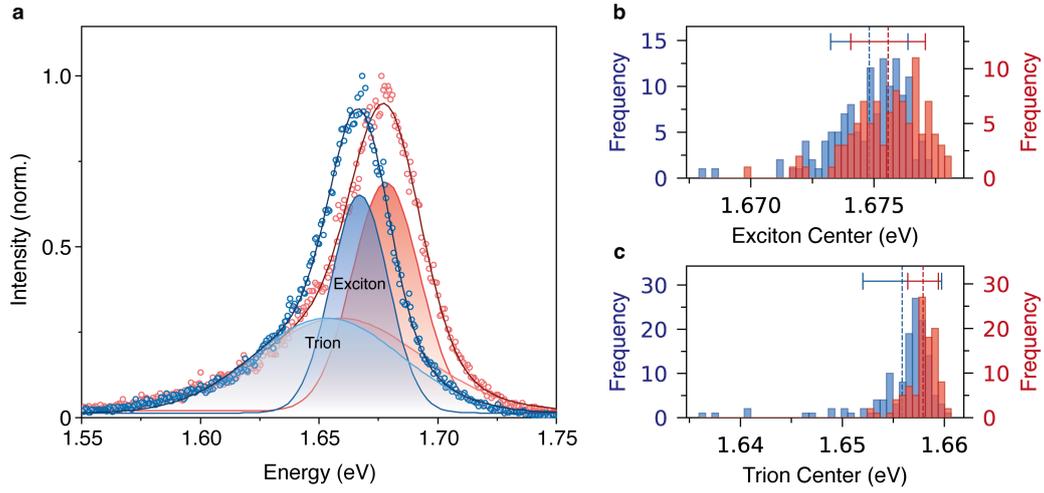

**Figure 3.** High-throughput spectral measurements of >200 WSe$_2$ monolayers exfoliated from different bulk crystals (dim and bright batch). (a) Selected room-temperature PL emission spectra of monolayers from the "dim" (blue) and "bright" (red) batches. Each spectrum is fitted by two Gaussian functions (shaded), where the solid lines show the sum. The trion and exciton emission peaks from the dim TMD monolayer are shifted to lower energies compared to the monolayer from the bright sample. (b,c) Distributions of the exciton (b) and trion (c) peak energies, obtained from the fitting procedure shown in panel a. The mean energy difference between the bright and dim sample is $\Delta E_\text{T} = -2.0$ meV and $\Delta E_\text{X} = -0.8$ meV. The vertical dashed lines and error bars in panels b and c indicate the mean and standard deviation of the distributions.

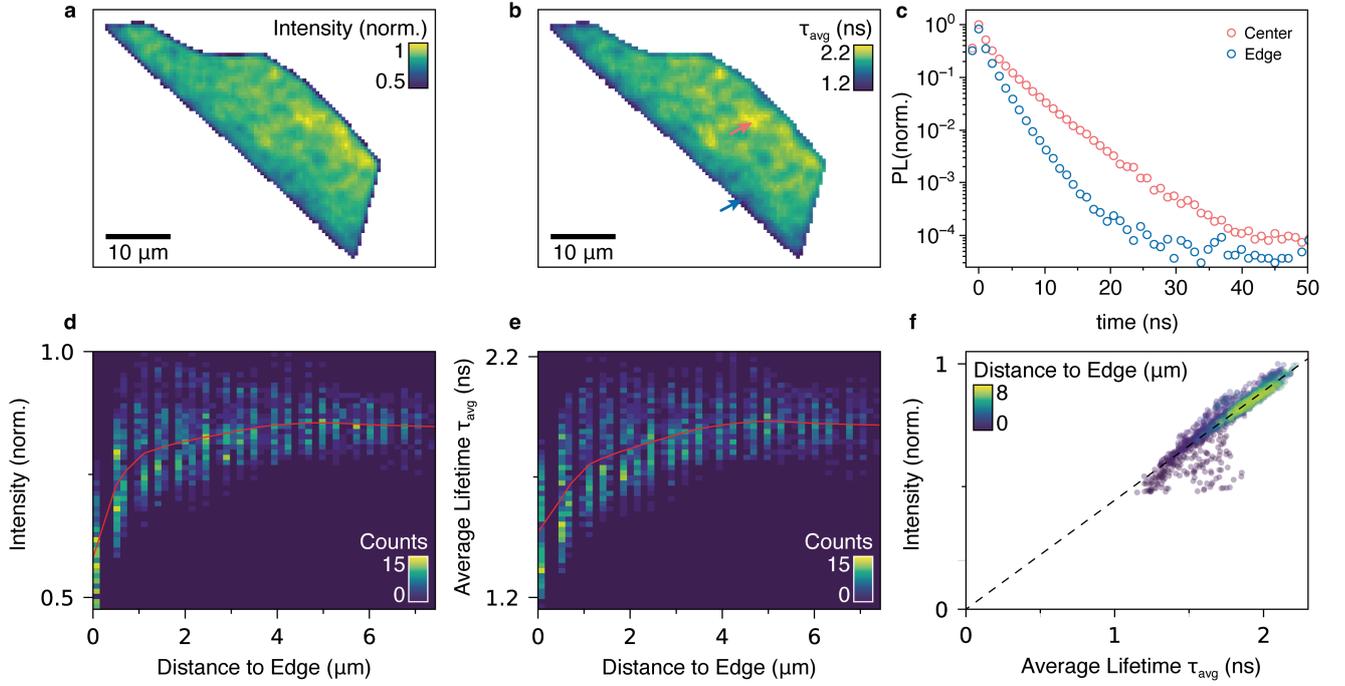

**Figure 4.** Spatial characterization of a single WSe$_2$ monolayer. (a) Intensity map of a single WSe$_2$ monolayer. To distinguish the monolayer from the background, we mask intensities $I < I_{max}/2$, where $I_{max}$ is the maximum measured intensity within the sample. We observe less bright emission from the edge of the monolayer. (b) Average-lifetime map for the same pixels selected in a. We observe shorter excited-state lifetimes at the edges of the monolayer. (c) PL decay curves from the center (red) and the edge (blue) of the monolayer. The exact positions within the monolayer are indicated by arrows in panel b. (d) Intensity versus distance to the edge plotted as a two-dimensional histogram. (e) Average lifetime versus distance to the edge plotted as a two-dimensional histogram. The red curves in d,e are LOWESS (locally weighted scatterplot smoothing) regressions. (f) Scatter plot of intensity versus average lifetime. Every plot point is color coded by the shortest distance from the pixel to the edge of the monolayer. Points further away from the edge are on average brighter and show a longer excited-state lifetime. The dashed line is a linear fit through the origin (see Table S1 in the Supporting Information).

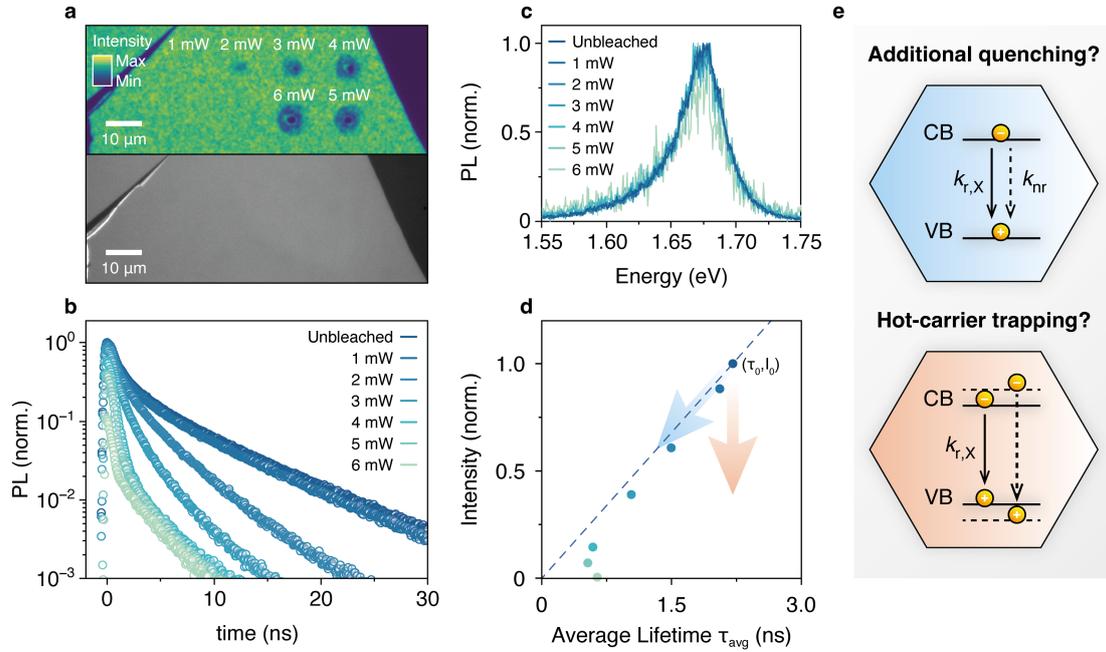

**Figure 5.** Controlled photobleaching of WSe$_2$. (a) A single monolayer under ambient conditions was exposed to increasing illumination from a 405-nm laser. Both the PL image (top) and BF image (bottom) were acquired after laser exposure. In the top image, each bleached spot is labeled by the excitation power used. (b) Decay curves from each bleached spot, together with a control measurement of an unbleached spot. The excited-state decay becomes faster the stronger a spot is bleached, indicating additional nonradiative decay pathways are induced by the 405-nm laser. (c) PL spectra of each bleached spot and an unbleached spot. All spectra are normalized to their maximum. (d) Scatter plot of intensity versus average lifetime for each bleached spot. The color code corresponds to the decay-curve measurements in b. The colored arrows indicate the expected change of intensity and average lifetime for additional quenching (blue) and hot-carrier trapping (red). The data is most consistent with additional nonradiative decay in competition with radiative decay (blue arrow). (e) Potential nonradiative decay pathways limiting PL brightness: (top) band-edge carrier trapping introduces an additional nonradiative decay pathway for the exciton, and (bottom) hot-carrier trapping limits the number of excitons that recombine at the band edge.

# Table of Contents Graphic

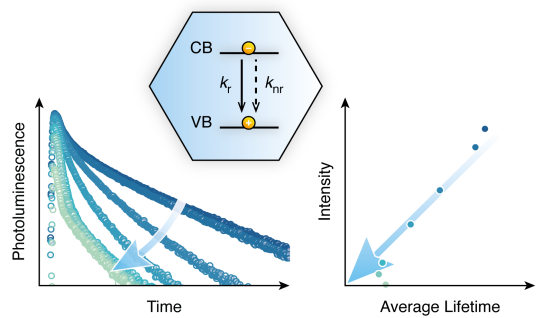

*Supporting Information for*

# Band-Edge Carrier Trapping Limits Light Emission in WSe$_2$

*Juri G. Crimmann,[†] Sander J. W. Vonk,[†] Yannik M. Glauser,[†]*
*Gabriel Nagamine,[†,§] and David J. Norris[*,†]*

[†]Optical Materials Engineering Laboratory, Department of Mechanical and Process Engineering, ETH Zurich, 8092 Zurich, Switzerland

[§]Institute of Applied Physics, University of Bern, 3012 Bern, Switzerland

**Corresponding Author**

*Email: dnorris@ethz.ch

## S1. Methods

**Sample Preparation.** WSe$_2$ bulk crystals (flux-grown) were purchased from 2D Semiconductors, and PDMS (PF-40X40-065-X4) from Gel-Pak. The samples were then prepared via mechanical exfoliation. In short, a small patch of PDMS was placed on a glass slide. A small amount of bulk WSe$_2$ was removed by pressing a piece of Scotch® Magic™ Tape onto a bulk crystal. Using multiple tapes, crystals were then thinned down. Next, a tape was slowly placed on the PDMS, slightly rubbed with a finger and a cotton swab. After 10 min, the tape was slowly removed. This process was done simultaneously for 10 PDMS patches. A more detailed step-by-step description is provided in Ref. S1.

**Photoluminescence Measurements.** The characterization of ~200 WSe$_2$ monolayers was done fully automatically. The workflow consisted of three steps. First, a sample was mapped (12 mm × 12 mm) by acquiring hundreds of PL images. Second, the images were analyzed to identify WSe$_2$ monolayers. Third, the stage moved to each monolayer individually, and PL spectra and lifetimes were acquired.

The mapping of the sample and the analysis is described in detail in Ref. S1. In short, we have used a white-light source as excitation, together with a 10× objective [Nikon, numerical aperture (NA) of 0.45, MRD00105]. With a set of optical filters (see Ref. S1), we collected only the PL of WSe$_2$ in a CMOS camera (Thorlabs DCU223M-GL). The stage was moved after each acquisition by two motorized actuators to the next position. After data collection, we analyzed the images using an intensity-thresholding procedure, resulting in a list of flakes, containing their position as well as geometric and PL properties, which were stored in a database.

For the characterization, a pulsed 405-nm laser (Picoquant, LDH-D-C-405) with the same optical filters and 10× objective was used. After passing a 700 nm longpass filter (Thorlabs, FELH0700), emission from a monolayer was directed towards a beamsplitter. One path was used for spectral measurements via a spectrometer (Andor, Shamrock 303i). The other path was used for time-resolved measurements with an avalanche photodiode (Excelitas, SPCM-AQRH-14-TR, single-photon timing resolution <250 ps).

The micro-PL map of a single WSe$_2$ monolayer was acquired as follows. The flake of interest was found using the mechanical actuators, as described above. We used a high-NA 50× objective (Nikon Plan Fluor, MUE13500, NA of 0.8) and piezo stage (Mad City Labs, NANO-LPS300) to measure a map with a step size of 500 nm. A PL spectrum and time-resolved measurement were acquired simultaneously at every point.

**Instrument Response Function.** We determined the instrument response function by directly measuring the laser light reflected from a silver mirror. The measurement is shown in Figure S7, where the asymmetric peak was fitted with a bi-Gaussian function. It is defined as

$$f(x) = \begin{cases} A + Be^{-\frac{1}{2}\left(\frac{x-x_c}{\sigma_1}\right)^2} & \text{if } x < x_c \\ A + Be^{-\frac{1}{2}\left(\frac{x-x_c}{\sigma_2}\right)^2} & \text{if } x \geq x_c \end{cases}$$

with constant background $A$, height $B$, center $x_c$, and standard deviations $\sigma_1$ and $\sigma_2$ of the left and right Gaussian, respectively. The width of the instrument response function is 0.59 ns, given by the sum of the individual widths $\sigma_1 = (0.127 \pm 0.003)$ ns and $\sigma_2 = (0.458 \pm 0.004)$ ns.

**Raman Measurements.** The measurements were done using an inVia Renishaw Raman microscope. We used a 50× objective (NA of 0.75), a 2400 lines/mm grating, a 514-nm excitation laser of less than 0.35 mW. The exposure time was 60 s.

**Atomic Force Microscopy Measurements.** We used an Oxford Instruments MFP-3D Origin AFM. Measurements were acquired using the tapping mode and AC160TS-R3 cantilevers.

**Average-Lifetime Analysis.** The average lifetime extracted from a time-resolved PL measurement was calculated from the fitting parameters (see Data-Fitting Procedures). The equation

$$\tau_{\text{avg}} = \frac{B_{\text{short}}\tau_{\text{short}} + B_{\text{long}}\tau_{\text{long}}}{B_{\text{short}} + B_{\text{long}}}$$

was used, containing the amplitudes $B_i$ and lifetimes $\tau_i$ of the short and long components.

**Data-Fitting Procedures.** PL spectra were fitted by two Gaussian functions and a constant background. A double-exponential decay function was used to fit time-resolved PL measurements:

$$f(t) = A + B_{\text{short}} e^{-\frac{t}{\tau_{\text{short}}}} + B_{\text{long}} e^{-\frac{t}{\tau_{\text{long}}}}$$

where $A$ is a constant background, and $B_i$ and $\tau_i$ are the amplitudes and lifetimes of the short and long components, respectively. In Figure S4, lifetime data was fitted by a double-exponential decay convoluted with a Gaussian instrument response function. We have performed power-dependent decay-curve measurements at variable number of excitons per pulse, $\bar{n}$. The average number of excitons is given by $\bar{n} = aI$, where $I$ is the laser intensity and $a$ a prefactor. The amplitudes of the long $A_{\text{long}}$ and short $A_{\text{short}}$ lifetime components were fitted by

$$A_{\text{long}}(\bar{n}) = A_{\text{long},\infty} P_{\geq 1}(\bar{n}) = A_{\text{long},\infty}[1 - P_0(\bar{n})]$$

$$A_{\text{short}}(\bar{n}) = A_{\text{short},\infty} P_{\geq 2}(\bar{n}) = A_{\text{long},\infty}[1 - P_0(\bar{n}) - P_1(\bar{n})]$$

where $A_{\text{long},\infty}/A_{\text{short},\infty}$ are the amplitudes at infinite excitation power, and $P_{\geq n}$ is the power-dependent probability of forming at least $n$ excitons. We find $P_{\geq n}$ from the Poissonian absorption statistics of exactly 0 ($P_0$) or 1 ($P_1$) absorption event(s), by complementarity. In the fitting procedure the amplitudes at infinite excitation power $A_{\text{long},\infty}/A_{\text{short},\infty}$, and the prefactor $a$ are free fitting parameters. Table S1 summarizes the linear fit used in Figures 4f of the main text.

## S2. SUPPLEMENTARY TABLE

| | $f(x) = mx$ |
|---|---|
| m (ns$^{-1}$) | 0.44 |
| R$^2$ | 0.86 |

**Table S1.** Overview of the fit function and fitting parameters used in Figure 4f of the main text.

# S3. SUPPLEMENTARY FIGURES

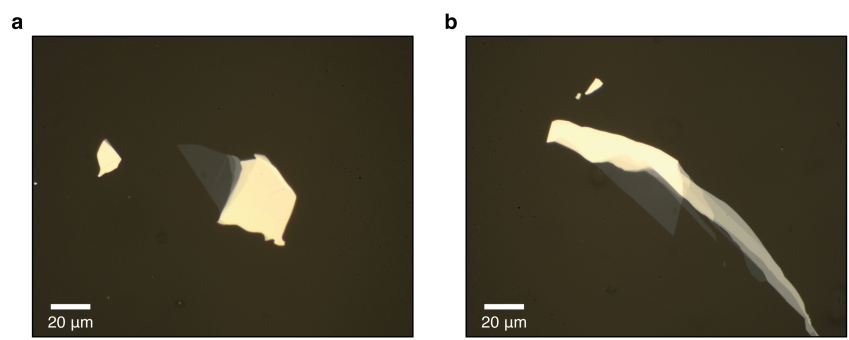

**Figure S1.** Brightfield images of the two WSe$_2$ monolayers on PDMS from Figure 1 in the main text.

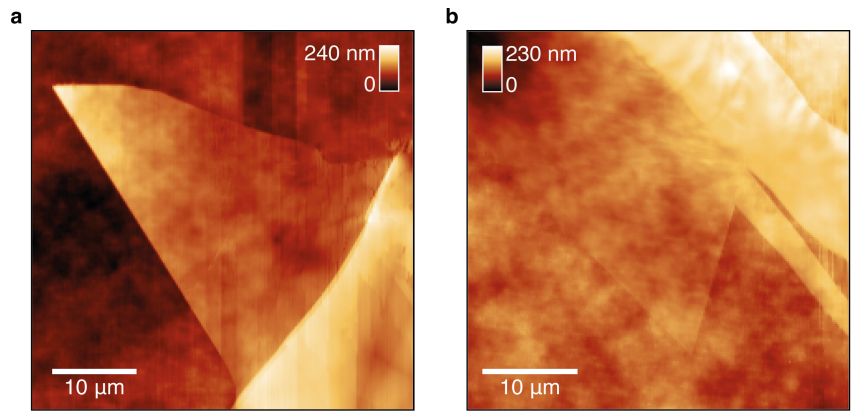

**Figure S2.** Atomic force microscopy images of the two WSe$_2$ monolayers on PDMS from Figure 1 in the main text.

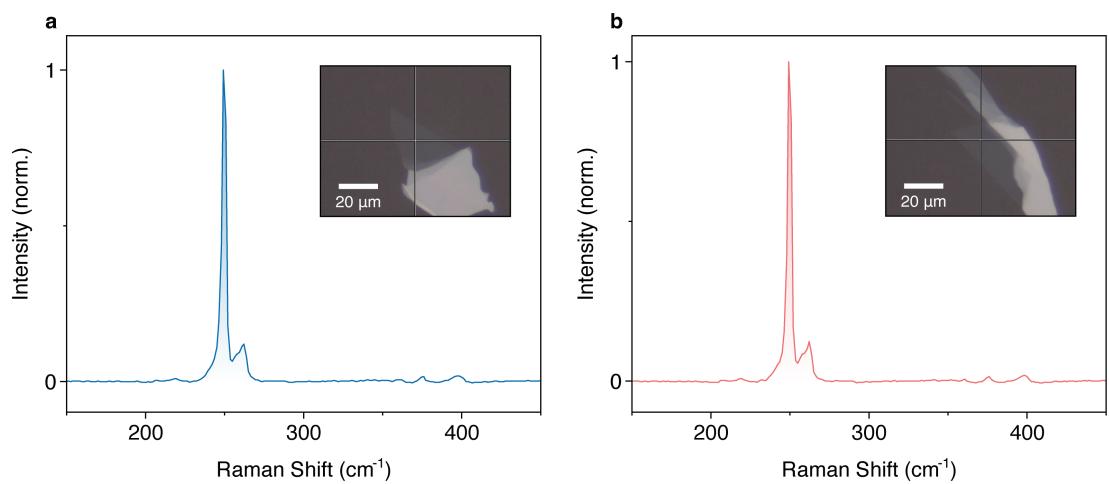

**Figure S3.** Raman spectra of the two WSe$_2$ monolayers on PDMS from Figure 1 in the main text. The insets show brightfield images of the flakes with a cross indicating where the Raman spectra were measured.

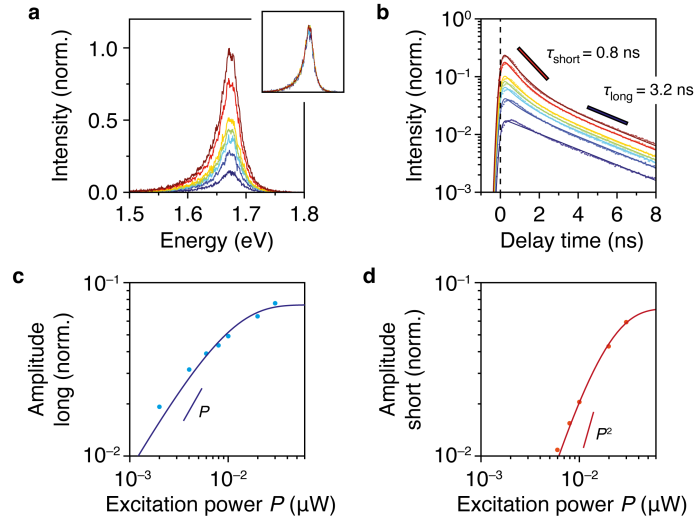

**Figure S4.** Power-dependent, room-temperature photoluminescence measurements of a single $WSe_2$ monolayer. (a) Spectra acquired with varying excitation power (brown – high, blue – low). The inset shows all spectra normalized. (b) Lifetime measurements of varying excitation power (brown – high, blue – low) with fits in solid lines. (c,d) The amplitude of the long and short-lifetime components as a function of excitation power on a log–log scale. Solid lines: fit functions given above in Section S1, which include the Poissonian absorption statistics on the amplitudes. For small excitation intensities, the long and short-lifetime components follow a linear/quadratic scaling, indicative of the formation of exciton/biexcitons(or charged biexcitons), respectively.

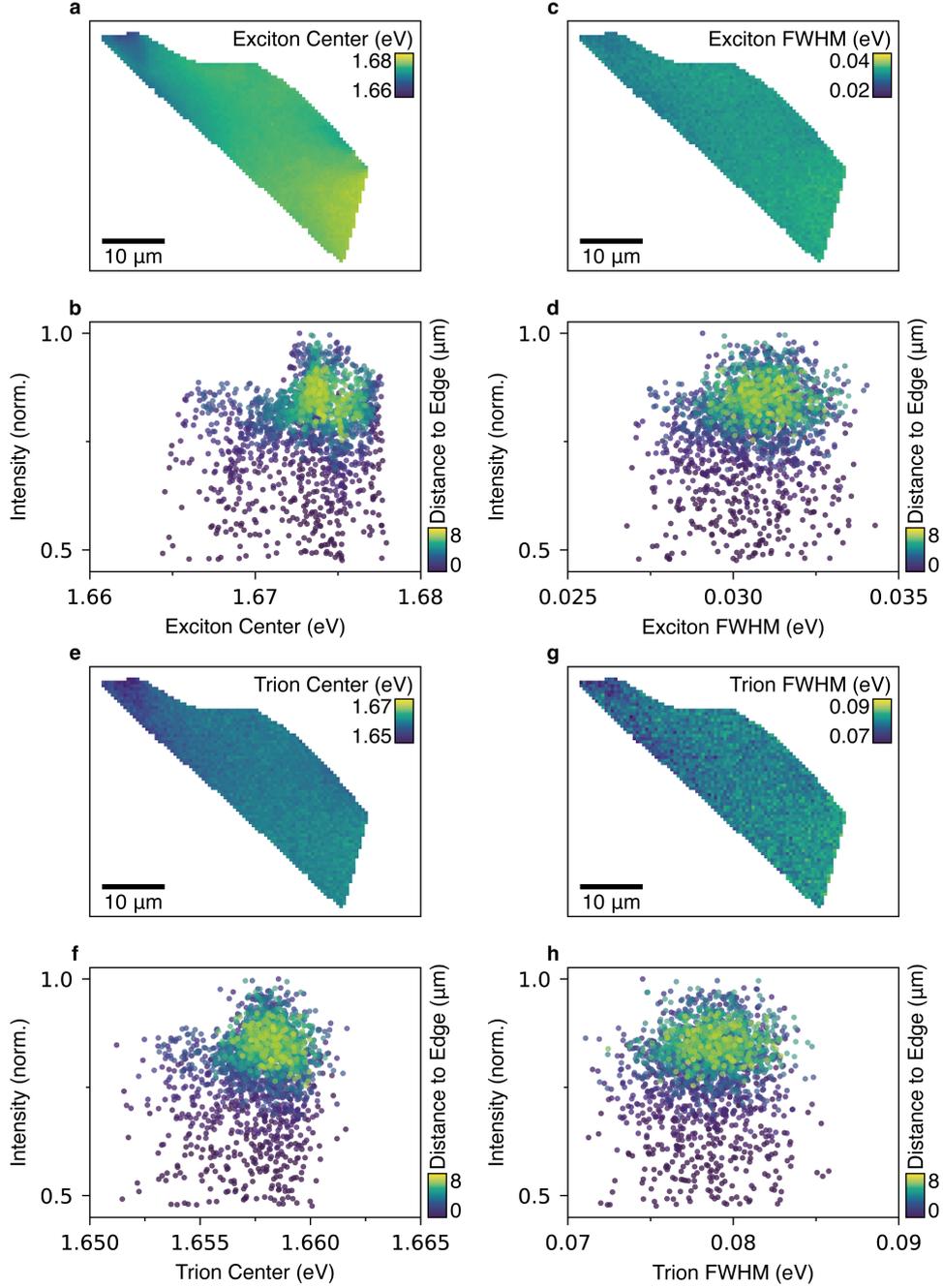

**Figure S5.** Room-temperature photoluminescence spectra maps of a WSe$_2$ monolayer. Spectra were acquired simultaneously with the lifetime data presented in Figure 4 of the main text. Spectra were analyzed like in Figure 3 of the main text by fitting two gaussians, one for the exciton and one for the trion emission. Exciton data is presented in panels (a–d), and trion data in panels (e–h). The peak emission energy and FWHM do not change significantly over a single flake.

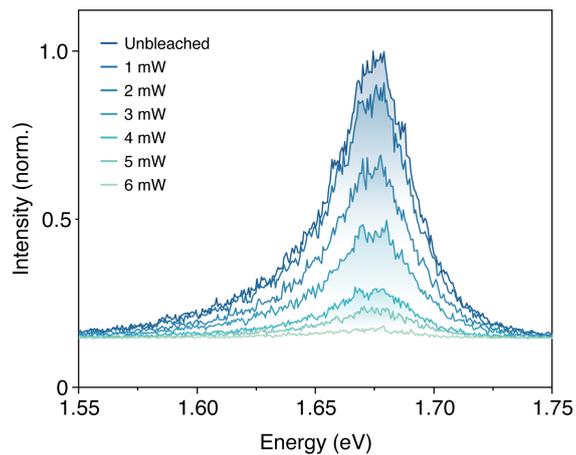

**Figure S6.** Room-temperature photoluminescence spectra of the photobleached WSe$_2$ monolayer of Figure 5 in the main text. The spectra were normalized to the maximum intensity of the unbleached spot, showing a gradual decrease in PL intensity with increased bleaching. The spectral position is unchanged upon photobleaching.

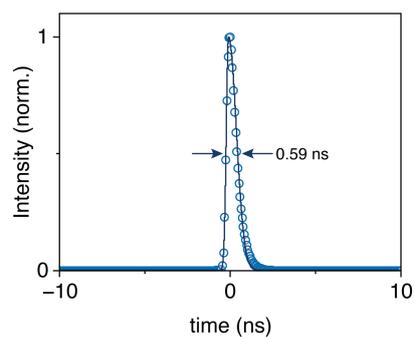

**Figure S7.** Measurement of the instrument response function. The data is fitted by a bi-Gaussian function (solid line), and the peak width is 0.59 ns. Details are provided in the Methods section of the Supporting Information.

## S4. SUPPLEMENTARY REFERENCES